\documentclass[aps,pre,preprint,groupedaddress,amssymb]{revtex4-2}
\usepackage{graphicx} 
\usepackage{natbib}
\usepackage{dcolumn}
\usepackage{bm}
\usepackage{amsmath}
\usepackage{epsfig}
\usepackage{color,hyperref}

\begin{document}
\title{Infinite ergodic theory and functional statistics of non-confined Feller process}

\author{Vicen\c c M\'endez}

\affiliation{Grup de F\'{\i}sica Estad\'{\i}stica, Departament de F\'{\i}sica. Facultat de Ci\`{e}ncies, Universitat Aut\`{o}noma de Barcelona, 08193 Barcelona, Spain}

\begin{abstract}
	We study additive observables of a non-confined Feller process characterized by a non-normalizable stationary probability density and return times with infinite mean.  The process is equivalent, after a deterministic rescaling, to a squared Bessel process, but we focus here on a question: the spatial structure of its local-time field. We show that the local time admits a factorization in the long time limit. Its spatial profile is governed by the non-normalizable stationary density, whereas its temporal fluctuations are controlled by a single Mittag-Leffler random amplitude. As a consequence, normalized spatial correlations of the local time converge to a universal constant independent of the two observation levels. Occupation times of finite intervals follow as corollaries and display Darling-Kac fluctuations. In contrast, non-integrable power observables exhibit self-similar squared-Bessel-type limits rather than Mittag-Leffler statistics. This spatial-field perspective provides a unifying framework for infinite ergodic theory under state-dependent multiplicative noise.
\end{abstract}

\maketitle

\section{Introduction}
\label{sec:intro}

Stochastic differential equations (SDEs) driven by  multiplicative noise constitute a fundamental paradigm across non-equilibrium statistical mechanics, condensed matter physics, quantitative finance, biophysics, and population biology \cite{Bouchaud1990, Cox1985, Gardin2009, Horsthemke1984}. Unlike standard Brownian motion with constant diffusion coefficients, multiplicative stochastic processes naturally exhibit non-Gaussian probability distributions, power-law tails, and complex dynamical phase transitions induced by the non-linear coupling between state variables and thermal or environmental fluctuations \cite{vanKampen2007, Schenzle1979}. A paradigmatic representative of this class is the Feller process—characterized by a linear drift term and a square-root multiplicative noise intensity \cite{Feller1951}. Originally introduced by Feller as a continuous-state limit of Galton-Watson branching processes, this model has achieved ubiquitous application, ranging from the Cox-Ingersoll-Ross (CIR) interest-rate model in mathematical finance \cite{Cox1985} to continuous-time random walk (CTRW) continuum limits in heterogeneous landscapes and anomalous diffusion in disordered media \cite{Barkai2014}.

In confined geometries or under strong restoring potentials, the competition between a inward drift and multiplicative noise stabilizes the system toward a well-defined, normalizable stationary probability distribution described by Boltzmann-Gibbs or Gamma-type densities \cite{Birkhoff1931, Risken1989}. In such ergodic settings, the classic Birkhoff Ergodic Theorem applies: long-time ensemble averages coincide with time averages calculated along a single trajectory for almost all realizations \cite{Birkhoff1931}. Conversely, in non-confined or weakly bound systems—corresponding in Feller process to drift parameter ranges $0 < \theta < 1$ with zero-flux boundary conditions at the origin—probability mass continuously escapes toward infinity. Consequently, the formal stationary solution of the Fokker-Planck equation yields an non-normalizable measure $\mathcal{I}(x)$, also known as infinite invariant density, whose spatial integral over the positive half-line diverges.

The physics of systems exhibiting an non-normalizable stationary density falls directly under the domain of \emph{Infinite Ergodic Theory} \cite{Aaronson1997, Korabel2009, Akimoto2010}. In this framework, standard ergodicity is fundamentally broken. Time-averaged observables do not converge to deterministic ensemble means even in the asymptotic limit $t \to \infty$; instead, they are random variables whose statistics vary from realization to realization \cite{Rebenshtok2007, Barkai2014, Deng2009}. For integrable physical observables $\mathcal{O}[x]$ satisfying $\int \mathcal{O}[x] dx < \infty$, the Aaronson-Darling-Kac theorem dictates that the time average $\overline{\mathcal{O}[X(t)]} = \frac{1}{t} \int_0^t U[X(t')] dt'$ scales sublinearly with observation time $t$, and its normalized fluctuations are governed by universal Mittag-Leffler limit distributions \cite{Aaronson1997, DarlingKac1957, He2008}.

While significant research has focused on single-point temporal statistics, first-passage time distributions, and spectral properties of infinite ergodic systems \cite{RevuzYor1999, Ke78, GoYo03, MaPe12}, the full spatial structure and multi-point correlation properties of local occupation fields remain largely unexplored. In particular, crucial physical questions remain open: How do fluctuations at different spatial locations $x_1$ and $x_2$ correlate over long observation times? Does the local-time field $L(x,t) = \int_0^t \delta(X(t') - x) dt'$ retain independent local spatial fluctuations, or does the recurrent nature of stochastic excursions induce global spatial coherence across the entire domain?

In this paper, we resolve these questions by establishing the exact spatiotemporal correlation structure, local-time field statistics, and residence-time distributions of non-confined Feller process ($0 < \theta < 1$). First, we demonstrate that the local-time field $L(x,t)$ undergoes an exact asymptotic spatiotemporal factorization into the product of the deterministic spatial invariant profile $\mathcal{I}(x)$ and a space-independent Mittag-Leffler random amplitude $\xi$. Second, as a direct measurable physical consequence of this factorization, we prove that the normalized two-point spatial correlation function flattens uniformly across space, converging to a universal, level-independent constant $2[\Gamma(2-\theta)]^2/\Gamma(3-2\theta)$. Third, we map this local-time field formulation onto residence times $T_a(t)$ and $T_a^+(t)$ for compact $[0,a]$ and non-compact $(a,\infty)$ spatial domains, establishing exact scaling collapse onto Mittag-Leffler statistics. Finally, we analyze the crossover to non-integrable power-law functionals $A_q(t) = \int_0^t X(t')^q dt'$, deriving the scaling form of its probability density function and the ergodicity breaking parameter.

The remainder of this paper is organized as follows. In Sec.~\ref{sec:feller}, we define the non-confined Feller diffusion model, write its Fokker-Planck operator, and detail the asymptotic properties of its exact transition density. In Sec.~\ref{sec:correlations}, we calculate the two-point cross-correlation function of the local-time field, proving spatiotemporal factorization and the existence of a universal spatial correlation plateau. In Sec.~\ref{sec:residence}, we analyze the residence-time statistics in compact and non-compact domains, deriving scaling laws and probability density functions. In Sec.~\ref{sec:nonintegrable}, we examine non-integrable power observables, obtaining the ergodicity breaking parameter and the scaling form of its probability density function. Finally, in Sec.~\ref{sec:conclusions}, we summarize our conclusions and discuss physical implications.
    
	\section{The Feller process}
	\label{sec:feller}
	
	\subsection{Confined Feller process}
	
	The general Feller process $Y(t)$ is a special class of diffusion process with linear drift and linear diffusion coefficient vanishing at the origin. The time evolution of the process is thus governed by the SDE
	\begin{equation}
		dY(t) = [\beta - \alpha Y(t)]\,dt + \sqrt{2D_0 Y(t)}\,dW(t), 
		\label{eq:feller_confined}
	\end{equation}
	where $Y(0) = y_0 \geq 0$. Here, $\alpha>0$ is the mean-reversion strength, $\beta > 0$ is a constant drift parameter, $D_0 > 0$ sets the multiplicative noise intensity, and $W(t)$ is the standard Wiener process, namely, a Gaussian process with zero mean, unit variance, and correlation function $\left\langle W(t_{1})W(t_{2})\right\rangle =\min(t_{1},t_{2})$. In what follows, all stochastic differentials are interpreted in the It\^o sense. 
	Equation~\eqref{eq:feller_confined} can be interpreted as a Langevin equation for a Brownian particle driven by multiplicative noise $\sqrt{2D_0 Y(t)}$ and moving in an effective potential given by
	\begin{equation}
		V(Y) = -\beta Y + \frac{\alpha}{2} Y^2.
		\label{eq:potential}
	\end{equation}
	The corresponding deterministic force, $F(Y) = -V'(Y) = \beta - \alpha Y$, captures the competition between two distinct mechanisms: the linear component of the potential ($-\beta Y$) provides a constant outward drift $+\beta$ that prevents the particle from collapsing at the origin $Y=0$, whereas the quadratic harmonic term ($\alpha Y^2/2$) confines the dynamics within a bounded spatial region.
	
	It is convenient to scale time and space (recalling $\alpha>0$) as: $t'=\alpha t$ and $X(t')=\alpha Y(t)/D_0$, so that the SDE reads
	\begin{equation}
		dX(t') = [\theta - X(t')]\,dt' + \sqrt{2 X(t')}\,dW(t'), 
		\label{eq:feller_confined2}
	\end{equation}
	where 
	\begin{equation}
		\theta = \frac{\beta}{D_0}
	\end{equation}
	is called the saturation or normal level to which the Feller process $X(t')$ is attracted. Unless otherwise stated, we drop the prime symbol on the time variable.
	
	The evolution of the probability density $P(x,t)$ is governed by the Fokker-Planck equation
	\begin{equation}
		\frac{\partial P(x,t)}{\partial t} = -\frac{\partial J(x,t)}{\partial x},
	\end{equation}
	with probability flux
	\begin{equation}
		J(x,t) = (\theta - x) P(x,t) - \frac{\partial}{\partial x}\left[ x P(x,t) \right].
		\label{eq:flux_confined}
	\end{equation}
	Because $\theta > 0$, a process initialized at a positive value cannot cross into the negative domain, remaining non-negative for all times. Consequently, for the Feller process, $x=0$ acts as a reflecting boundary. In the stationary regime ($\partial P/\partial t = 0$), continuity requires a spatially uniform flux $J(x,t) = J_0$. Enforcing a zero-flux condition at $x=0$ sets $J_0 = 0$, yielding
	\begin{equation}
		(\theta - x) P_{\text{st}}(x) - \frac{d}{dx}\left[ x P_{\text{st}}(x) \right] = 0.
		\label{ode}
	\end{equation}
	Solving Eq.~\eqref{ode} gives $P_{\text{st}}(x)=\mathcal{N}x^{\theta-1}e^{-x}$. Normalization requires $\int_0^\infty P_{\text{st}}(x)dx=\mathcal{N}\Gamma (\theta)=1$, yielding the Gamma distribution
	\begin{equation}
		P_{\text{st}}(x)=\frac{1}{\Gamma(\theta)}x^{\theta-1}e^{-x},
        \label{gd}
	\end{equation}
	which is well-defined and normalizable on $(0,\infty)$ provided $\theta > 0$.
	
	\subsection{Non-confined Feller process}
	
	Setting the confining parameter to $\alpha = 0$, Eq.~\eqref{eq:feller_confined} reduces to the non-confined Feller process
	\begin{equation}
		dY(t) = \beta\,dt + \sqrt{2D_0Y(t)}\,dW(t), 
		\label{eq:felleru}
	\end{equation}
	with $Y(0)=y_0\geq 0$. This SDE describes a squared Bessel process of dimension $\beta$ \cite{RevuzYor1999}. Rescaling time and space yields
	\begin{equation}
		dX(t') = \theta\,dt' + \sqrt{2 X(t')}\,dW(t'),
		\label{eq:feller}
	\end{equation}
	with probability flux
	\begin{equation}
		J(x,t) = \theta P(x,t) - \frac{\partial}{\partial x}\left[ x P(x,t) \right].
		\label{eq:flux_intro}
	\end{equation}
	Enforcing zero probability flux at $x=0$ yields the stationary spatial profile 
    \begin{eqnarray}
        P_{\text{st}}(x) = \mathcal{N}x^{\theta-1}.
        \label{pst2}
    \end{eqnarray}
    The integral over $(0,\infty)$ diverges when $0 < \theta < 1$, rendering $P_{\text{st}}(x)$ an \textit{non-normalizable} stationary density (or infinite invariant density). Physically, the probability mass persistently disperses toward $+\infty$.
	
	\section{Ergodic theory}
	\label{sec:ergodic_theory}
	
	\subsection{Standard ergodic theory and ergodicity breaking}
	
	We first outline the fundamental observables and metrics within ergodic theory. Consider a general observable $\mathcal{O}[X(t)]$, defined as a functional over a trajectory realization $\{X(t'); 0 \le t' \le t\}$. Because the path $X(t)$ is stochastic, the observable $\mathcal{O}[X(t')]$ fluctuates between independent realizations of the process.
	
	Let $P(x,t|x_0)$ be the propagator denoting the probability density that $X(t)=x$ given $X(0)=x_0$. If $\mathcal{O}[x]$ is integrable with respect to $P(x,t|x_0)$, its ensemble average at time $t$ reads
	\begin{equation}
		\left\langle \mathcal{O}[X(t)] \right\rangle = \int_{-\infty}^{\infty} \mathcal{O}[x] P(x,t|x_0) \, dx.
		\label{eq:ensemble_avg}
	\end{equation}
	Conversely, the time average accumulated along a single trajectory of duration $t$ is defined as
	\begin{equation}
		\ensuremath{\overline{\mathcal{O}[X(t)]}} = \frac{1}{t} \int_0^t \mathcal{O}[X(t')] \, dt'.
		\label{tav}
	\end{equation}
In classical ergodic theory on a finite measure space, the propagator $P(x,t|x_0)$ converges asymptotically to a normalizable invariant density $P_{\text{st}}(x)$, i.e., $P_{\text{st}}(x)=\lim_{t\to\infty}P(x,t|x_0)$. If $\mathcal{O}[x]$ is integrable with respect to $P_{\text{st}}(x)$, then Birkhoff's ergodic theorem establishes that $\mathcal{O}[x]$ is ergodic if its long-time average converges to the stationary ensemble mean
    \begin{eqnarray}
\lim_{t\to\infty}\overline{\mathcal{O}}=\lim_{t\to\infty}\frac{1}{t}\int^{t}_{0}\mathcal{O}[X(t')]\,dt'=\langle\mathcal{O}\rangle_{\text{st}}=\int^{\infty}_{-\infty}\mathcal{O}[x]P_{\mathrm{st}}(x)\,dx.
    \end{eqnarray}
    
    In standard ergodic systems bound by a normalizable stationary density $P_{\text{st}}(x)$, individual trajectory averages converge strictly to a deterministic limit. As a consequence, trajectory-to-trajectory fluctuations vanish asymptotically, and the limiting probability density function (PDF) of the time average, $P(\overline{\mathcal{O}},t)$, collapses to a Dirac delta function
	\begin{equation}
		\lim_{t \to \infty} P(\overline{\mathcal{O}}, t) = \delta\left( \overline{\mathcal{O}} - \left\langle \overline{\mathcal{O}} \right\rangle \right).
		\label{lpdf}
	\end{equation}
	
	For an ergodic observable, the variance of the time average $\text{Var}(\overline{\mathcal{O}})$ decays to zero as $t \to \infty$. Conversely, in non-ergodic regimes, $\overline{\mathcal{O}}$ is a random variable even in the limit $t \to \infty$, exhibiting persistent variance across realizations. To quantify departures from deterministic convergence and measure sample-to-sample fluctuations between individual trajectories, one introduces the \textit{ergodicity breaking parameter} $\textrm{EB}$
	\begin{equation}
		\textrm{EB} = \lim_{t \to \infty} \frac{\text{Var}(\overline{\mathcal{O}})}{\left\langle \overline{\mathcal{O}} \right\rangle^2} = \lim_{t \to \infty} \frac{\left\langle \overline{\mathcal{O}}^2 \right\rangle - \left\langle \overline{\mathcal{O}} \right\rangle^2}{\left\langle \overline{\mathcal{O}} \right\rangle^2}.
		\label{EB}
	\end{equation}
	For ergodic observables, $\textrm{EB} = 0$, indicating that a single trajectory uniformly covers the available space and renders the time average deterministic. Conversely, $\textrm{EB} > 0$ characterizes weak ergodicity breaking (WEB), where trajectory time averages retain sample-dependent randomness governed by a non-trivial PDF at arbitrarily long times.
	
	As a concrete example, consider the additive functional 
	\begin{eqnarray}
		Z(t) = \int_0^t U[X(t')] dt'
		\label{Z}
	\end{eqnarray}
and the observable $\mathcal{O}[X(t)] = U[X(t)]$. Then the time average can be written in terms of the functional as $\overline{\mathcal{O}[X(t)]} = Z(t)/t$. The first two moments of the time average are thus related to $Z(t)$ via
	\begin{equation}
		\left\langle \overline{\mathcal{O}} \right\rangle = \frac{\left\langle Z(t) \right\rangle}{t}, \qquad \left\langle \overline{\mathcal{O}}^2 \right\rangle = \frac{\left\langle Z(t)^2 \right\rangle}{t^2}.
		\label{o}
	\end{equation}
	Substituting these expressions into Eq.~\eqref{EB} provides the ergodicity breaking parameter for $U[X(t')]$ over $t' \in [0,t]$
	\begin{equation}
		\textrm{EB}(Z) = \lim_{t \to \infty} \frac{\left\langle Z(t)^2 \right\rangle}{\left\langle Z(t) \right\rangle^2} - 1.
		\label{EB2}
	\end{equation}
	If the system has a normalizable invariant density density $P_{\text{st}}(x)$ and the observable $U[x]$ is integrable with respect to $P_{\text{st}}(x)$, then from the Birkhoff’s theorem the PDF of $Z(t)/t$ converges to a Dirac delta function centered at its ensemble stationary value
    \begin{eqnarray}
        \lim_{t\to\infty}P\left(\frac{Z(t)}{t}\right)=\delta\left(\frac{Z(t)}{t}-\left\langle Z\right\rangle _{\mathrm{st}}\right)
        \label{ldelta}
    \end{eqnarray}
where
$$
\left\langle Z\right\rangle _{\mathrm{st}}=\int_{\mathcal{S}}U[x]P_{\text{st}}(x)dx.
$$
If $X(t)$ follows the SDE of the confined Feller process given in \eqref{eq:feller_confined2} then $P_{\text{st}}(x)$ is given by \eqref{gd}. Then, the limiting PDF of $Z(t)/t$ is given by 
\eqref{ldelta} with
$$
\left\langle Z\right\rangle _{\mathrm{st}}=\frac{1}{\Gamma(\theta)}\int_{0}^{\infty}U[x]x^{\theta-1}e^{-x}dx.
$$

\subsection{Infinite ergodic theory and invariant density}

 Infinite ergodic systems are characterized by an non-normalizable invariant measure. Importantly, we assume that the trajectories are recurrent: they return to the relevant region infinitely often with probability one. 

Infinite ergodic theory establishes that $P(x,t|x_0)$ does not decay featurelessly. For a fixed spatial coordinate $x$, it maintains an invariant spatial profile at long times by compensating the temporal decay with a scaling factor $t^{1-\gamma}$, where $\gamma \in (0,1)$ is the return (persistence) exponent, determined by the tail of the return time distribution, which is assumed to decay as $t^{-1-\gamma}$ as $t \to \infty$. The infinite invariant density $\mathcal{I}(x)$ is non-normalized (i.e., $\int_{\mathcal{S}}\mathcal{I}(x)dx=\infty$ where $\mathcal{S}$ is the spatial domain) and is formally defined by
\begin{equation}
	 \mathcal{I}(x) = \lim_{t \to \infty} \mathcal{C} t^{1-\gamma} P(x,t|x_0),
	\label{eq:infinite_density_def}
\end{equation}
where $\mathcal{C}$ is a model-dependent constant. Although $\mathcal{I}(x)$ satisfies the stationary Fokker-Planck equation under zero probability flux conditions, it fails to constitute a normalizable probability measure over $\mathcal{S}$. However, this infinite invariant density governs long-time statistics, controls rare events, and determines measurable observables. 

Because the integral of $\mathcal{I}(x)$ over $\mathcal{S}$ diverges, standard Birkhoff ergodic theory breaks down: the time average of any local integrable observable $\mathcal{O}[X(t)] $ vanishes. Assuming $\mathcal{O}[X(t)]$ is integrable with respect to $\mathcal{I}(x)$, its long-time ensemble average becomes
\begin{eqnarray}
	\left\langle \mathcal{O}\left[X(t)\right]\right\rangle &=&\int_{\mathcal{S}}\mathcal{O}[x]P(x,t|x_{0})dx\nonumber\\
	&\underset{t\to\infty}{\approx}&\frac{t^{\gamma-1}}{\mathcal{C}}\int_{\mathcal{S}}\mathcal{O}[x]\mathcal{I}(x)dx.
	\label{mo}
\end{eqnarray}
Since $\gamma<1$, $\left\langle \mathcal{O}\left[X(t)\right]\right\rangle \to 0 \text{ as } t \to \infty$. This relation indicates that despite the non-normalizable nature of $\mathcal{I}(x)$, statistical averages can still be evaluated by integrating over this density, analogous to averaging with normalized invariant measures. Next, consider the ensemble average of the time average
\begin{eqnarray}
\left\langle \overline{\mathcal{O}\left[X(t)\right]}\right\rangle &=&\frac{1}{t}\int^{t}_{0}dt'\int_{\mathcal{S}}\mathcal{O}\left[x\right]P(x,t'|x_{0})dx\nonumber\\
&=&\frac{1}{t}\int^{t}_{0}dt'\left\langle \mathcal{O}\left[X(t')\right]\right\rangle. 
\end{eqnarray}
Therefore, using \eqref{mo},
\begin{eqnarray}
\left\langle \overline{\mathcal{O}\left[X(t)\right]}\right\rangle \underset{t\to\infty}{\approx}\frac{1}{\gamma}\left\langle \mathcal{O}\left[X(t)\right]\right\rangle, 
\label{property}
\end{eqnarray}
which directly connects the ensemble average of the time average to the instantaneous ensemble mean. Non-trivial long-time statistics are recovered within infinite ergodic theory. 

For example, we can apply property \eqref{property} to stochastic functionals. The mean value of the functional defined in \eqref{Z} grows universally as $t^\gamma$ if $\int_{\mathcal{S}}U[x]\mathcal{I}(x)dx<\infty$. Indeed, considering \eqref{o} and \eqref{mo},
\begin{eqnarray}
\left\langle Z(t)\right\rangle \underset{t\to\infty}{\approx}\frac{t^{\gamma}}{\gamma\mathcal{C}}\int_{\mathcal{S}}U[y]\mathcal{I}(y)dy,
\label{mzl}
\end{eqnarray}
which corresponds to a sublinear temporal growth provided that $\gamma<1$. 

Furthermore, if the underlying random process (here, the non-confined Feller process) is recurrent with a first-passage time PDF decaying as $t^{-1-\gamma}$ and possesses an infinite invariant density, then for any functional $Z(t)$ integrable with respect to $\mathcal{I}(x)$, the Darling-Kac theorem is satisfied. In consequence, the normalized random variable
\begin{eqnarray}
\xi=\lim_{t\to\infty}\frac{\gamma\overline{\mathcal{O}}}{\left\langle \mathcal{O}\right\rangle }=\lim_{t\to\infty}\frac{Z(t)}{\left\langle Z(t)\right\rangle }
\label{xi}
\end{eqnarray}
with unit mean, is distributed according to the Mittag-Leffler distribution 
\begin{eqnarray}
\mathcal{M}_\gamma(\xi)=\frac{\Gamma(1+\gamma)^{1/\gamma}}{\gamma\xi^{1+\frac{1}{\gamma}}}L_{\gamma}\left[\frac{\Gamma(1+\gamma)^{1/\gamma}}{\xi^{\frac{1}{\gamma}}}\right],
\label{mld}
\end{eqnarray}
where $L_\gamma (z)$ is the one-sided L\'evy density of order $\gamma$, defined through the inverse Laplace transform $L_{\gamma}(z)=\mathcal{L}^{-1}_{s\to z}\left[e^{-s^{\gamma}}\right]$ ($\mathcal{L}$ denotes Laplace transform). As a consequence, the EB parameter of $Z(t)$ is given by \eqref{EB2} and \eqref{xi} as
\begin{eqnarray}
	\mathrm{EB}=\left\langle \xi^{2}\right\rangle -1=\frac{2\Gamma\left(1+\gamma\right)^{2}}{\Gamma\left(1+2\gamma\right)}-1.
    \label{ebx}
	\end{eqnarray}
The parameter $\mathrm{EB}$ exhibits a strict universality: it is entirely independent of the specific choice of the local observable $U[x]$, provided that it is integrable with respect to $\mathcal{I}(x)$. It depends uniquely on the persistence exponent $\gamma$, which governs the asymptotic tail of the return-time distribution. 

For the confined Feller process, in the previous subsection we showed that the steady state is governed by a normalized Gamma distribution. In this case, any integrable observable with respect to the invariant density is ergodic and the limiting PDF of its time average is a Dirac delta function centered at its mean value.

Now consider the unconfined Feller process, whose stationary state profile is given in Eq.~\eqref{pst2}. It can be shown that this profile is actually an infinite invariant density. Solving the Fokker--Planck equation corresponding to \eqref{eq:feller} for the propagator $P(x,t|x_0)$ yields \cite{MasoliverPerello2012}
\begin{eqnarray}
P(x,t|x_{0})=\frac{1}{t}\left(\frac{x}{x_{0}}\right)^{\frac{\theta-1}{2}}e^{-\frac{x+x_{0}}{t}}I_{\theta-1}\left(\frac{2\sqrt{xx_{0}}}{t}\right),
    \label{prop2}
\end{eqnarray}
where $I_\nu(z)$ is the modified Bessel function of the first kind of order $\nu$. For fixed $x$ and taking $t\to\infty$,
\begin{eqnarray}
P(x,t|x_{0})\underset{t\to\infty}{\approx}\frac{x^{\theta-1}}{t^{\theta}\Gamma\left(\theta\right)},
	\label{eq:kernel_asymp}
\end{eqnarray}
where we have used the approximation $I_\nu(z) \sim \frac{(z/2)^\nu}{\Gamma(\nu+1)}$ for small argument. Note that this solution has the structure of \eqref{eq:infinite_density_def} with 
\begin{eqnarray}
    \mathcal{I}(x)=x^{\theta-1}
    \label{I}
\end{eqnarray}
and $\mathcal{C}=\Gamma(\theta)$, provided $\theta$ is related to the persistence exponent $\gamma$ via $\theta = 1-\gamma$. We must confirm that the persistence exponent indeed equals $1-\theta$. The first-passage or return time to a given level is known to decay as $t^{-2+\theta}$ as $t \to \infty$  \cite{Ke78,GoYo03,MaPe12}, confirming $\theta = 1-\gamma$. 

This non-normalizable stationary profile naturally divides additive observables into two classes: local observables integrable with respect to $\mathcal{I}(x)$ that follow Darling-Kac Mittag-Leffler statistics, and non-integrable observables governed by distinct self-similar scaling limits.
Within this classification, our primary object of interest is the local-time field
\begin{equation}
	L(x,t)=\int_0^t\delta(X(t')-x)\,dt',
	\label{eq:localtime}
\end{equation}
defined as the spatial density of the occupation time spent by the trajectory at or in the immediate neighborhood of a specific value $x$ up to time $t$. The spatial correlation structure and field dynamics of $L(x,t)$ under an infinite invariant density remain unexplored. We demonstrate that the local-time field exhibits an asymptotic spatial-temporal factorization
\begin{equation}
	L(x,t)\underset{t\to\infty}{\approx}\mathcal{I}(x)\Phi(t),
	\label{eq:factorization_intro}
\end{equation}
where $\Phi (t)$ is a spatially uniform, global random amplitude. We also show that all normalized local times converge to the same Mittag-Leffler random variable, giving rise to universal, level-independent spatial correlations.

	\section{Local time: first moment}
	\label{sec:localtime}
To compute the first moment of the local time, we directly average its definition. From Eq.~\eqref{eq:localtime}, we obtain
\begin{eqnarray}
    \left\langle L(x,t)\right\rangle &=&\int^{t}_{0}\left\langle \delta\left(X(t')-x\right)\right\rangle dt'=\int^{t}_{0}dt'\int^{\infty}_{0}\delta(x-y)P(y,t'|x_{0})dy\nonumber\\
    &=&\int^{t}_{0}P(x,t'|x_{0})dt'.
    \label{mL}
\end{eqnarray}
Obtaining an explicit expression for the mean local time requires integrating the propagator. While Eq.~\eqref{prop2} yields the exact result, we focus on the long-time behavior by substituting Eq.~\eqref{eq:kernel_asymp} into Eq.~\eqref{mL}:
\begin{eqnarray}
    \left\langle L(x,t)\right\rangle \underset{t\to\infty}{\approx}\frac{\left(t/x\right)^{1-\theta}}{(1-\theta)\Gamma(\theta)}.
    \label{mL2}
\end{eqnarray}
Alternatively, this result can be derived from the infinite ergodic theory using Eq.~\eqref{mzl}, given that the local time is integrable with respect to the invariant density. Setting $\gamma = 1-\theta$, $\mathcal{C}=\Gamma(\theta)$, $\mathcal{I}(y)=y^{\theta-1}$ from Eq.~\eqref{I}, and $U[y]=\delta(y-x)$, Eq.~\eqref{mzl} reproduces Eq.~\eqref{mL2}.

\begin{figure}[h!]
    \centering
    \includegraphics[width=0.65\linewidth]{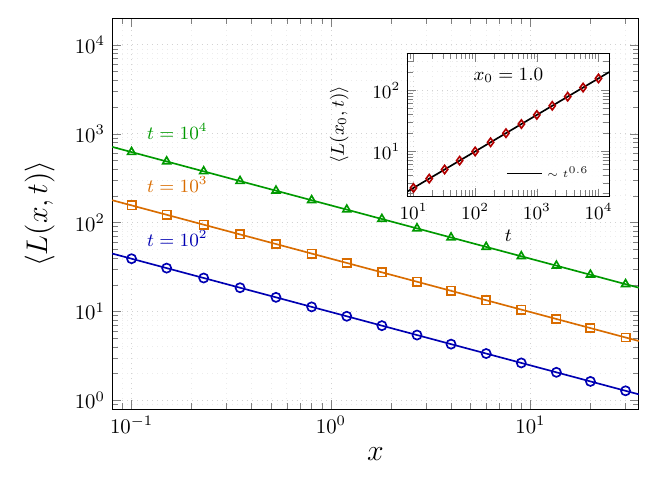}
\caption{Main panel: Log-log plot of the mean local time $\langle L(x,t) \rangle$ as a function of position $x$ for three different observation times ($t = 100, 1000, 10000$). Inset: Time dependence of $\langle L(x_0,t) \rangle$ at a fixed spatial position $x_0 = 1.0$, displaying a clear power-law growth behavior $\sim t^{0.6}$. In both panels $\theta=0.4$. Symbols correspond to simulations and solid curves are the theoretical values obtained from Eq. \eqref{mL2}. }
\label{fig:spatial_temporal_scaling}
    \label{fig1}
\end{figure}

In Figure \ref{fig1} we show the behavior of the mean local time with $x$ and $t$. In the main panel we plot the mean local time versus $x$ for three different values of the observation time $t$ in a log-log scale. In the inset  we plot the mean local time versus $t$ for $x=1$. The agreement between the theoretical results depicted with solid lines provided by Eq. \eqref{mL2} and the numerical simulations (symbols) is excellent. The details of the numerical simulations ca be found in the appendix. 
	
\section{Spatial correlations of local time}
	\label{sec:correlations}
	
We now examine local-time fluctuations at two distinct observation levels, $x_1$ and $x_2$, by computing the two-point correlation $\left\langle L(x_{1},t)L(x_{2},t)\right\rangle$.
Using definition~\eqref{eq:localtime}, we write
\begin{eqnarray}
   \left\langle L(x_{2},t)L(x_{1},t)\right\rangle =\int^{t}_{0}dt_{2}'\int^{t}_{0}dt_{1}'\left\langle \delta\left(X(t_{2}')-x_{2}\right)\delta\left(X(t_{1}')-x_{1}\right)\right\rangle.
   \label{cc1}
\end{eqnarray}
The integrand represents the expectation of a product of two Dirac delta functions, which can be evaluated using the joint density $P(x_2,t_2;x_1,t_1|x_0,0)$. This density specifies the probability of finding the process at level $x_1$ at time $t_1$ and at level $x_2$ at time $t_2$, given $X(0)=x_0$. The integrand is then
\begin{eqnarray*}
    \left\langle \delta\left(X(t_{2}')-x_{2}\right)\delta\left(X(t_{1}')-x_{1}\right)\right\rangle =\int^{\infty}_{0}dx_{2}'\delta(x_{2}'-x_{2})\int^{\infty}_{0}dx_{1}'\delta(x_{1}'-x_{1})P(x_{2}',t_{2}';x_{1}',t_{1}'|x_{0},0).
    \label{ccf}
\end{eqnarray*}
The joint density can be factored as
\begin{eqnarray}
    P(x_{2}',t_{2};x_{1}',t_{1}|x_{0},0)=P(x_{2}',t_{2}|x_{1}',t_{1};x_{0},0)P(x_{1}',t_{1}|x_{0},0)
    \label{p1}
\end{eqnarray}
by definition of conditional probability. Furthermore, applying the Markov property gives
\begin{eqnarray}
    P(x_{2}',t_{2}|x_{1}',t_{1};x_{0},0)=P(x_{2}',t_{2}|x_{1}',t_{1}).
    \label{p2}
\end{eqnarray}
Inserting \eqref{p2} into \eqref{p1}, the two-point joint density becomes
\begin{eqnarray}
    P(x_{2}',t_{2};x_{1}',t_{1}|x_{0},0)=P(x_{2}',t_{2}|x_{1}',t_{1})P(x_{1}',t_{1}|x_{0},0).
    \label{mark}
\end{eqnarray}
To exploit the Laplace transform of a convolution, we shift the time variables. Because the Feller process is time-homogeneous, $P(x_{2}',t_{2}|x_{1}',t_{1})=P(x_{2}',t_{2}-t_{1}|x_{1},0)$, allowing Eq.~\eqref{mark} to be rewritten as
\begin{eqnarray}
    P(x_{2}',t_{2};x_{1}',t_{1}|x_{0},0)=P(x_{2}',t_{2}-t_{1}|x_{1},0)P(x_{1}',t_{1}|x_{0},0).
    \label{mark2}
\end{eqnarray}
Substituting Eq.~\eqref{mark2} into Eq.~\eqref{p1} and integrating yields
\begin{eqnarray}
    \left\langle \delta\left(X(t_{2}')-x_{2}\right)\delta\left(X(t_{1}')-x_{1}\right)\right\rangle =\left\{ \begin{array}{cc}
P(x_{2},t_{2}'-t_{1}'|x_{1},0)P(x_{1},t_{1}'|x_{0},0), & t_{2}'>t_{1}'\\
P(x_{1},t_{1}'-t_{2}'|x_{2},0)P(x_{2},t_{2}'|x_{0},0), & t_{2}'<t_{1}'.
\end{array}\right.
\label{cc2}
\end{eqnarray}
Inserting Eq.~\eqref{cc2} into Eq.~\eqref{cc1} and splitting the integration domain via $\int^{t}_{0}dt_{1}'\int^{t}_{0}dt_{2}'=\int^{t}_{0}dt_{1}'\int^{t}_{t_{1}'}dt_{2}'+\int^{t}_{0}dt_{2}'\int^{t}_{t_{2}'}dt_{1}'$, we obtain
\begin{eqnarray}
	 \left\langle L(x_{2},t)L(x_{1},t)\right\rangle &=&\int^{t}_{0}dt_{1}'P(x_{1},t_{1}'|x_{0},0)\int^{t}_{t_{1}'}dt_{2}'P(x_{2},t_{2}'-t_{1}'|x_{1},0)\nonumber\\
     &+&\int^{t}_{0}dt_{2}'P(x_{2},t_{2}'|x_{0},0)\int^{t}_{t_{2}'}dt_{1}'P(x_{1},t_{1}'-t_{2}'|x_{2},0).
\end{eqnarray}
Introducing the relative time variables $\tau=t_2'-t_1'$ and $\tau'=t_1'-t_2'$ into the first and second integrals, respectively, gives
\begin{eqnarray}
    \left\langle L(x_{2},t)L(x_{1},t)\right\rangle &=&\int^{t}_{0}dt_{1}'P(x_{1},t_{1}'|x_{0},0)\int^{t-t_{1}'}_{0}P(x_{2},\tau|x_{1},0)d\tau\nonumber\\
    &+&\int^{t}_{0}dt_{2}'P(x_{2},t_{2}'|x_{0},0)\int^{t-t_{2}'}_{0}P(x_{1},\tau'|x_{2},0)d\tau'.
    \label{ll}
\end{eqnarray}
Applying the Laplace transform $\mathcal{L}[f(t)]\equiv\widetilde{f}(s)=\int^{\infty}_{0}f(t)e^{-st}dt$ to Eq.~\eqref{ll} via the convolution theorem yields
\begin{eqnarray}
    \mathcal{L}\left[\left\langle L(x_{2},t)L(x_{1},t)\right\rangle \right]=\frac{1}{s}\left[\widetilde{P}(x_{2},s|x_{1})\widetilde{P}(x_{1},s|x_{0})+\widetilde{P}(x_{1},s|x_{2})\widetilde{P}(x_{2},s|x_{0})\right].
    \label{lll}
\end{eqnarray}
Here, $\widetilde{P}(x,s|y)$ solves the Laplace-transformed Fokker--Planck equation associated with Eq.~\eqref{eq:feller}. In the time domain, the propagator satisfies
\begin{eqnarray}
   \frac{\partial P(x,t|y)}{\partial t}=-\theta\frac{\partial P(x,t|y)}{\partial x}+\frac{\partial^{2}}{\partial x^{2}}\left[xP(x,t|y)\right] 
   \label{FP}
\end{eqnarray}
subject to the initial condition $P(x,0|y)=\delta (x-y)$. Taking the Laplace transform to Eq. \eqref{FP} converts this partial differential equation into the ordinary differential equation
\begin{eqnarray}
    x\frac{d^{2}\widetilde{P}(x,s|y)}{dx^{2}}+(2-\theta)\frac{d\widetilde{P}(x,s|y)}{dx}-s\widetilde{P}(x,s|y)=-\delta(x-y),
\end{eqnarray}
which must be solved under the reflecting (zero-flux) boundary conditions at $x=0$ and $\widetilde{P}(x\to +\infty,s|y)=0$. The delta function splits the domain into $x \in [0,y]$ and $x \in [y,+\infty)$. The solution satisfying the boundary requirements is
$$
\widetilde{P}(x,s|y)=\left\{ \begin{array}{cc}
c_{1}x^{\frac{\theta-1}{2}}I_{\theta-1}\left(2\sqrt{sx}\right), & 0<x<y\\
c_{2}x^{\frac{\theta-1}{2}}K_{\theta-1}\left(2\sqrt{sx}\right), & x>y
\end{array}\right..
$$
The coefficients $c_1$ and $c_2$ are determined by matching continuity at $x=y$ and enforcing the jump condition on the derivative, $\left.\frac{d\widetilde{P}}{dx}\right|_{x=y^+} - \left.\frac{d\widetilde{P}}{dx}\right|_{x=y^-} = -1/y$. Using the identities 
$$\frac{d}{dz}\left[z^{\nu}I_{\nu}(z)\right]=z^{\nu}I_{\nu-1}(z),\quad \frac{d}{dz}\left[z^{\nu}K_{\nu}(z)\right]=-z^{\nu}K_{\nu-1}(z),$$
we obtain
\begin{eqnarray}
    \widetilde{P}(x,s|y)=2\left(\frac{y}{x}\right)^{\frac{1-\theta}{2}}\cdot\left\{ \begin{array}{cc}
K_{\theta-1}\left(2\sqrt{sy}\right)I_{\theta-1}\left(2\sqrt{sx}\right), & 0<x<y\\
K_{\theta-1}\left(2\sqrt{sx}\right)I_{\theta-1}\left(2\sqrt{sy}\right), & x>y
\end{array}\right..
\label{Pxy}
\end{eqnarray}
Substituting this result into Eq.~\eqref{lll} and taking the small-$s$ (long-time) limit using the asymptotic expansions
$$
K_{\theta-1}\left(2\sqrt{sy}\right)\approx\frac{\Gamma(1-\theta)}{2}(sy)^{\frac{\theta-1}{2}},\quad I_{\theta-1}\left(2\sqrt{sy}\right)\approx\frac{(sy)^{\frac{\theta-1}{2}}}{\Gamma(\theta)}
$$
yields, after simplification
$$
\mathcal{L}\left[\left\langle L(x_{2},t)L(x_{1},t)\right\rangle \right]\underset{s\to 0}{\approx}\frac{2\Gamma(1-\theta)^{2}}{\Gamma(\theta)^{2}s^{3-2\theta}(x_{1}x_{2})^{1-\theta}},
$$
which upon Laplace inversion gives
\begin{eqnarray}
\left\langle L(x_{2},t)L(x_{1},t)\right\rangle \underset{t\to\infty}{\approx}\frac{2\Gamma(1-\theta)^{2}}{\Gamma(\theta)^{2}\Gamma(3-2\theta)}\left(\frac{t^{2}}{x_{1}x_{2}}\right)^{1-\theta}.
\label{lla}
\end{eqnarray}
The ergodicity breaking parameter for the local time follows directly from Eqs.~\eqref{mL2} and \eqref{lla}. Inserting these expressions into Eq.~\eqref{EB2} yields
\begin{eqnarray}
    \text{EB}(L)=\lim_{t\to\infty}\frac{\left\langle L(x,t)^{2}\right\rangle }{\left\langle L(x,t)\right\rangle ^{2}}-1=\frac{2\Gamma\left(2-\theta\right)^{2}}{\Gamma\left(3-2\theta\right)}-1,
    \label{eb3}
\end{eqnarray}
which is remarkably independent of the level $x$. This behavior is expected because the criteria for the Darling--Kac theorem are satisfied. Consequently, the normalized variable $\xi = \lim_{t\to \infty} L(x,t)/ \left\langle L(x,t)\right\rangle$ follows the Mittag-Leffler distribution $\mathcal{M}_\gamma(\xi)$ [Eq.~\eqref{mld}], with its EB parameter given by Eq.~\eqref{ebx} for $\gamma=1-\theta$.

Crucially, we must verify that $L(x,t)/\left\langle L(x,t)\right\rangle$ becomes independent of the spatial coordinate $x$ as $t \to \infty$. This spatial independence unveils a key property of local-time correlations. We define the normalized two-point correlation function as
\begin{eqnarray}
  R(x_{2},x_{1},t)\equiv\frac{\left\langle L(x_{2},t)L(x_{1},t)\right\rangle }{\left\langle L(x_{2},t)\right\rangle \left\langle L(x_{1},t)\right\rangle }.  
  \label{R}
\end{eqnarray}
From \eqref{mL2} and \eqref{lla}, it follows that
\begin{eqnarray}
    \lim_{t\to\infty}R(x_{2},x_{1},t)=\frac{2\Gamma\left(2-\theta\right)^{2}}{\Gamma\left(3-2\theta\right)}.
    \label{rlt}
\end{eqnarray}

The striking feature of Eq.~\eqref{rlt} is its complete independence of $x_1$ and $x_2$, demonstrating that local-time fluctuations across different levels $x$ are asymptotically governed by a single global random amplitude $\Xi(t)$. In the following section, we formally prove that in the long-time limit, the local field $L(x,t)$ factorizes into the product of $\mathcal{I}(x)$ and $\Xi(t)$. 
The exact agreement between the limiting correlation $R(x_2, x_1, t)$ [Eq.~\eqref{rlt}] and the second moment of a Mittag-Leffler variable [Eq.~\eqref{ebx}] directly manifests weak ergodicity breaking as governed by the Darling--Kac theorem. 

\begin{figure}[h!]
    \centering
    \includegraphics[width=0.65\linewidth]{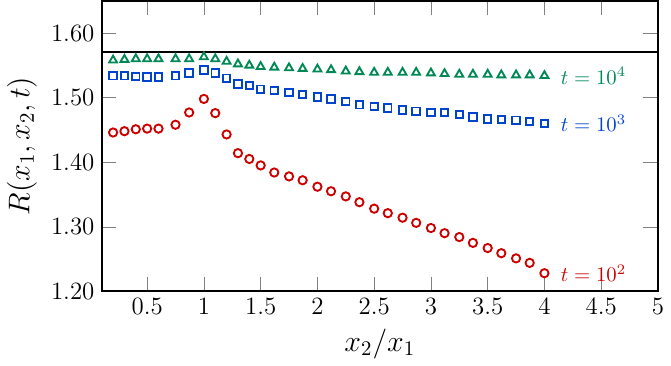}
\caption{Spatial correlation function $R(x_1, x_2, t)$ as a function of the ratio $x_2/x_1$ for $x_1 = 1.0$ and $\theta = 0.5$. Symbols denote simulation results evaluated at observation times $t = 10^2$ (orange circles), $t = 10^3$ (purple squares), and $t = 10^4$ (green triangles). The solid black horizontal line represents the theoretical universal constant $\lim_{t\to\infty}R(x_{2},x_{1},t) = 2[\Gamma(2-\theta)]^2 / \Gamma(3-2\theta) \approx 1.571$. }
\label{fig:spatial_temporal_scaling}
    \label{fig2}
\end{figure}
In Figure 2 we plot the normalized two-point correlation function versus the ratio $x_2/x_1$. For the numerical simulations we have made use of \eqref{R} for three fixed values of the observation time. As time increases ($t \to \infty$), the spatial dependence flattens uniformly across all ratios $x_2/x_1$, demonstrating the asymptotic emergence of rigid spatial coherence and a convergence to the uniform value given by Eq. \eqref{rlt}. 
    
\section{Asymptotic factorization of the local-time field}
	\label{sec:factorization}
In this section, we formally demonstrate that the ratio $L(x,t)/\left\langle L(x,t)\right\rangle$ becomes spatially uniform as $t \to \infty$, and that Eq.~\eqref{rlt} implies the exact spatiotemporal factorization of the local-time field $L(x,t)$.
We define the normalized local-time field $\xi(x,t)$ as
\begin{equation}
\xi(x,t)\equiv\frac{L(x,t)}{\left\langle L(x,t)\right\rangle }.
\label{eq:normalized_field}
\end{equation}
By construction, $\left\langle \xi(x,t)\right\rangle = 1$ for all positions $x$.
The two-point local-time correlation $R(x_2,x_1,t)$ from Eq.~\eqref{R} is nothing but the autocorrelation of this rescaled field
\begin{equation}
R(x_{2},x_{1},t)=\left\langle \xi(x_{2},t)\xi(x_{1},t)\right\rangle.
\end{equation}

As $t \to \infty$, $R(x_2,x_1,t)$ converges to a constant $C$ independent of $x_1$ and $x_2$ [Eq.~\eqref{rlt}], yielding the following identities for any pair of observation levels $x_1, x_2$
\begin{align}
\lim_{t\to\infty}\left\langle \xi(x_{2},t)^{2}\right\rangle =\lim_{t\to\infty}\left\langle \xi(x_{1},t)^{2}\right\rangle =\lim_{t\to\infty}\left\langle \xi(x_{2},t)\xi(x_{1},t)\right\rangle =C.
\label{eq:crosscorr}
\end{align}

To evaluate the spatial variability of $\xi(x,t)$, we compute the mean-squared difference between the normalized local time at positions $x_1$ and $x_2$
\begin{equation}
\left\langle \left(\xi(x_{2},t)-\xi(x_{1},t)\right)^{2}\right\rangle =\left\langle \xi(x_{2},t)^{2}\right\rangle +\left\langle \xi(x_{1},t)^{2}\right\rangle -2\left\langle \xi(x_{2},t) \xi(x_{1},t)\right\rangle. 
\label{expansion}
\end{equation}

Substituting Eq.~\eqref{eq:crosscorr} into expansion~\eqref{expansion} yields
\begin{equation}
\lim_{t\to\infty}\left\langle \left(\xi(x_{2},t)-\xi(x_{1},t)\right)^{2}\right\rangle =0.
\end{equation}

Because the expectation of a non-negative random variable vanishes if and only if the variable itself is zero almost surely, we conclude that
\begin{equation}
\lim_{t\to\infty}\xi(x_{2},t)=\lim_{t\to\infty}\xi(x_{1},t)
\end{equation}
almost surely for all $x_1$ and $x_2$.
Thus, $\xi(x,t)$ becomes spatially uniform in the long-time limit, collapsing to a single space-independent global stochastic amplitude $\Xi(t) \equiv \lim_{t\to\infty}\xi(x,t)$. Via Eq.~\eqref{eq:normalized_field}, this allows us to write
\begin{equation}
L(x,t)\underset{t\to\infty}{\approx}\left\langle L(x,t)\right\rangle \Xi(t),
\label{las}
\end{equation}
where $\left\langle L(x,t)\right\rangle$ is given by Eq.~\eqref{mL2}. Furthermore, $\Xi(t)$ is time-independent because $\lim_{t\to\infty}\xi(x,t)=\zeta$ is a stationary Mittag--Leffler random variable. Consequently, the local time simplifies to
\begin{eqnarray}
    L(x,t)\underset{t\to\infty}{\approx}\frac{\mathcal{I}(x)t^{1-\theta}}{(1-\theta)\Gamma(\theta)}\zeta,
    \label{lf}
\end{eqnarray}
where $\zeta$ is the Mittag-Leffler random variable. The global random amplitude introduced in Eq. \eqref{eq:factorization_intro} is thus
\begin{eqnarray}
    \Phi(t)\equiv\frac{t^{1-\theta}}{(1-\theta)\Gamma(\theta)}\zeta .
    \label{Xi}
\end{eqnarray}
Consequently, the PDF of $L(x,t)$ is obtained via a standard change of variables. 
The probability density that $L(x,t) = L$ at long times is
\begin{eqnarray}
    P(L,x,t)\underset{t\to\infty}{\approx}\frac{1}{\left\langle L(x,t)\right\rangle }\mathcal{M}_{1-\theta}\left(\frac{L}{\left\langle L(x,t)\right\rangle }\right),
    \label{plt}
\end{eqnarray}
where $\mathcal{M}_{1-\theta}(\xi)$ is the Mittag-Leffler function defined in \eqref{mld} with $\gamma=1-\theta$. Using the asymptotic behavior of the one-sided L\'evy distribution for small and large arguments \cite{Ba01,Penson2010}
$$
L_{\alpha}(z)\underset{z\to0}{\approx}\frac{B}{z^{\sigma}}e^{-\kappa z^{-\rho}},\quad L_{\alpha}(z)\underset{z\to\infty}{\approx}\frac{\alpha}{\Gamma(1-\alpha)z^{1+\alpha}},
$$
where 
$$
B=\sqrt{\frac{\alpha^{\frac{1}{1-\alpha}}}{2\pi(1-\alpha)}},\;\sigma=\frac{2-\alpha}{2-2\alpha},\;\kappa=(1-\alpha)\alpha^{\frac{\alpha}{1-\alpha}},\;\rho=\frac{\alpha}{1-\alpha},
$$
we obtain the asymptotic tail behavior
\begin{eqnarray}
P(L,x,t)\underset{t\to\infty}{\approx}\left\{ \begin{array}{cc}
\frac{1}{\Gamma(2-\theta)\Gamma(\theta)\left\langle L(x,t)\right\rangle }, & L\ll\left\langle L(x,t)\right\rangle \\
A_{\theta}\left(\frac{L^{1-2\theta}}{\left\langle L(x,t)\right\rangle }\right)^{\frac{1}{2\theta}}\exp\left[-\lambda_{\theta}\left(\frac{L}{\left\langle L(x,t)\right\rangle }\right)^{\frac{1}{\theta}}\right], & L\gg\left\langle L(x,t)\right\rangle 
\end{array}\right.
\label{pdfap}
\end{eqnarray}
with
$$
A_{\theta}=\frac{(1-\theta)^{\frac{1-3\theta}{2\theta}}}{\sqrt{2\pi\theta}[\Gamma(2-\theta)]^{\frac{1}{2\theta}}},\quad \lambda_{\theta}=\frac{\theta(1-\theta)^{\frac{1-\theta}{\theta}}}{[\Gamma(2-\theta)]^{1/\theta}}.
$$

\begin{figure}[htbp]
\centering
\includegraphics[width=0.65\linewidth]{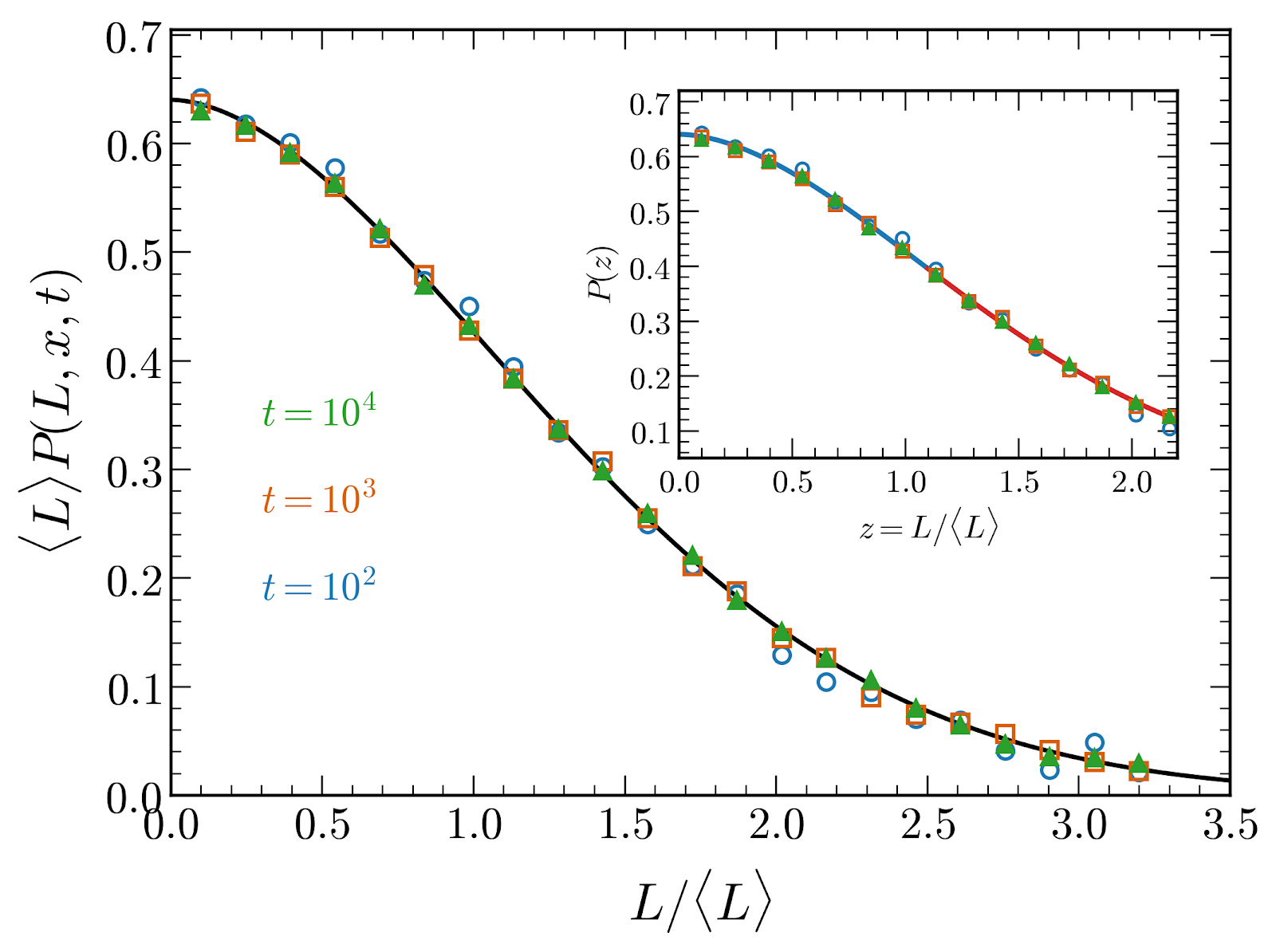}
\caption{Rescaled probability density $\langle L \rangle P(L,x,t)$ versus $L/\langle L(x,t) \rangle$. Symbols represent numerical simulations at different time steps ($t = 10^2, 10^3, 10^4$), displaying an exact data collapse onto the theoretical curve (solid black line) given by Eq. \eqref{mld}. Inset: Behavior of $P(z)$ as a function of $z = L/\langle L(x,t) \rangle$, highlighting asymptotic fits for small $z$ (solid blue line) and large $z$ (solid red line).}
\label{fig3}
\end{figure}

In Figure 3 we plot the rescaled PDF $\left\langle L(x,t)\right\rangle P(L,x,t)$ for large values of the observation time. To compute the solid lines we employ Eqs. \eqref{plt}, \eqref{mld} and \eqref{mL2}. For different values of $t$ the numerical data collapse on the theoretical curve given by the Mittag-Leffler distribution. In the inset we check the approximations given in Eq. \eqref{pdfap} (solid lines) for the inner part and the tail of the PDF. The numerical data is the same as for the main panel. The agreement between our theoretical results and numerical simulations is again excellent.

	\section{Occupation times}
	\label{sec:residence}
Let $T_a(t)$ and $T_a^+(t)$ denote the occupation times up to time $t$ in the intervals $[0,a]$ and $[a,\infty)$, respectively. These times satisfy $T_a(t)+T_a^+(t)=t$, with
\begin{eqnarray}
    T_{a}(t)=\int_{0}^{t}\theta\left(a-X(t')\right)dt',
\end{eqnarray}
where $\theta(\cdot)$ is the Heaviside step function. The occupation time $T_a(t)$ directly relates to the local time via
\begin{eqnarray}
    \int_{0}^{a}L(x,t)dx=\int_{0}^{t}dt'\int_{0}^{a}\delta\left(X(t')-x\right)dx=\int_{0}^{t}dt'\theta\left(a-X(t')\right)=T_{a}(t).
    \label{tal}
\end{eqnarray}
Substituting Eq.~\eqref{lf} into Eq.~\eqref{tal} expresses $T_a(t)$ in terms of the Mittag-Leffler variable
\begin{eqnarray}
    T_{a}(t)\underset{t\to\infty}{\approx}\frac{t^{1-\theta}}{(1-\theta)\Gamma(\theta)}\zeta\int_{0}^{a}\mathcal{I}(x)dx=\frac{a^{\theta}t^{1-\theta}}{\theta(1-\theta)\Gamma(\theta)}\zeta.
\end{eqnarray}
Averaging this expression yields the first moment
\begin{eqnarray}
    \left\langle T_{a}(t)\right\rangle \underset{t\to\infty}{\approx}\frac{a^{\theta}t^{1-\theta}}{\theta(1-\theta)\Gamma(\theta)}.
    \label{mta}
\end{eqnarray}
The normalized variable $\lim_{t\to\infty}T_a(t)/\left\langle T_{a}(t)\right\rangle = \zeta$ aligns with Eq.~\eqref{xi} and follows the Mittag-Leffler distribution. Consequently, the PDF of $T_a(t)$ is
\begin{eqnarray}
    P(T_{a},t)\underset{t\to\infty}{\approx}\frac{1}{\left\langle T_{a}(t)\right\rangle }\mathcal{M}_{1-\theta}\left(\frac{T_{a}}{\left\langle T_{a}(t)\right\rangle }\right),
    \label{pta1}
\end{eqnarray}
where $\left\langle T_{a}(t)\right\rangle$ is given in \eqref{mta}. This PDF shares the structure of the local-time PDF upon replacing $L$ with $T_a$. This equivalence is expected because both functionals fulfill the conditions of the Darling-Kac theorem. The ergodicity breaking parameters coincide, $\mathrm{EB}(T_a) = \mathrm{EB}(L)$, and are given by Eq.~\eqref{ebx} with $\gamma = 1-\theta$.

Similarly, the PDF of the complementary occupation time $T_a^+(t)$ is
\begin{eqnarray}
P(T_{a}^{+},t)\underset{t\to\infty}{\approx}\frac{1}{\left\langle T_{a}(t)\right\rangle }\mathcal{M}_{1-\theta}\left(\frac{t-T_{a}^{+}}{\left\langle T_{a}(t)\right\rangle }\right),
\label{ptamas}
\end{eqnarray}
and its ergodicity breaking parameter is
\begin{eqnarray}
    \mathrm{EB}(T^{+}_{a})=\lim_{t\to\infty}\frac{\left\langle T^{+}_{a}(t)^{2}\right\rangle }{\left\langle T^{+}_{a}(t)\right\rangle ^{2}}-1=\mathrm{EB}(T_{a})\lim_{t\to\infty}\left(\frac{\left\langle T_{a}(t)\right\rangle }{t-\left\langle T_{a}(t)\right\rangle }\right)^{2}=0,
\end{eqnarray}
where we used Eq.~\eqref{mta}. Thus, as the observation time increases, the complementary occupation time dominates and becomes asymptotically equivalent to the total time $t$.

Additionally, we can compute the PDFs of the occupation fractions $\eta = T_a/t$ and $\eta^+ = T_a^+/t$. The PDF of $\eta$ reads
\begin{eqnarray}
    P(\eta,t)\underset{t\to\infty}{\approx}\frac{1}{\epsilon(t)}\mathcal{M}_{1-\theta}\left(\frac{\eta}{\epsilon(t)}\right),
\end{eqnarray}
where $\epsilon(t)=a^\theta /[\theta(1-\theta)\Gamma(\theta)t^\theta]$. Recalling that the Laplace transform of the Mittag-Leffler distribution yields the Mittag-Leffler function \cite{Go20}
$$
\int_{0}^{\infty}e^{-su}\mathcal{M}_{\alpha}(u)du=E_{\alpha}\left(-s\right),
$$
we have $\tilde{P}(u,t) = E_{1-\theta}\left[-u\epsilon(t)\right]$. Since $\epsilon(t) \sim t^{-\theta} \to 0$ as $t \to \infty$ and $E_{1-\theta}(0) = 1$, $\tilde{P}(u,t) \to 1$. Inverting the Laplace transform yields
\begin{eqnarray}
 \lim_{t\to\infty}P(T_{a}/t)=\delta(T_{a}),
 \label{delta1}
\end{eqnarray}
and since $T_a+T_a^+=t$,
\begin{eqnarray}
  \lim_{t\to\infty}P(T_{a}^{+}/t)=\delta(T_{a}^{+}-t).  
  \label{delta2}
\end{eqnarray}
For every fixed threshold $a>0$, the non-confined Feller process spends an asymptotically unit fraction of the observation time above $a$, although the absolute occupation time of the compact interval $(0,a)$ diverges sublinearly and retains nontrivial Mittag-Leffler fluctuations.

\begin{figure}[htbp]
\centering
\includegraphics[width=0.65\linewidth]{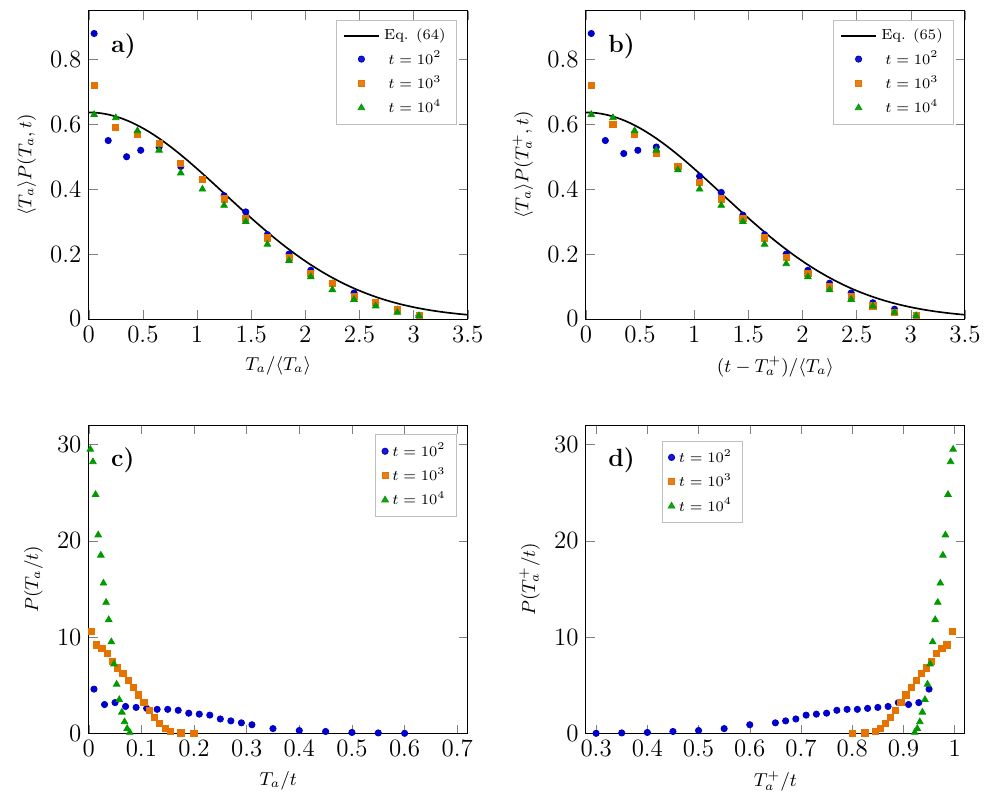}
\caption{
(a)~Rescaled probability density of $T_a(t)$, $\langle T_a \rangle P(T_a, t)$, as a function of $T_a / \langle T_a \rangle$, compared with the theoretical Mittag-Leffler scaling function $M_{1-\theta}(z)$ given by Eq. \eqref{pta1}. 
(b)~Equivalent scaling collapse for $T_a^+(t)$, $\langle T_a \rangle P(T_a^+, t)$, versus $(t - T_a^+) / \langle T_a \rangle$, along with the theoretical prediction of Eq.~\eqref{ptamas}. 
(c)~PDF of the time fraction $T_a / t$. 
(d)~PDF of the time fraction $T_a^+ / t$ . 
In all panels, symbols represent simulation results for $t = 10^2$ (blue circles), $t = 10^3$ (orange squares), and $t = 10^4$ (green triangles), while solid black lines denote the theoretical curves.}
\label{fig4}
\end{figure}

In Figure \ref{fig4} (panels a) and b)) we plot the rescaled PDF of $T_a$ and $T_a^+$, $\left\langle T_{a}(t)\right\rangle P(T_a,t)$ and $\left\langle T_{a}(t)\right\rangle P(T_a^+,t)$, to check the validity  of Eqs. \eqref{pta1} and \eqref{ptamas}. We see how numerical simulation data collapse for different large values of the observation time, on the theoretical curves. In panels c) and d) we plot the PDF of $T_a/t$ and $T_a^+/t$. We note that as time increases from $t=10^2$ to $t=10^4$ we data is peaked around the origin (panel c)) or towards 1 (panel d)) in agreement with the theoretical predictions given in Eqs. \eqref{delta1} and \eqref{delta2}.

\section{Nonintegrable power observables}
	\label{sec:nonintegrable}
\begin{figure}[htbp]
\centering
\includegraphics[width=0.5\linewidth]{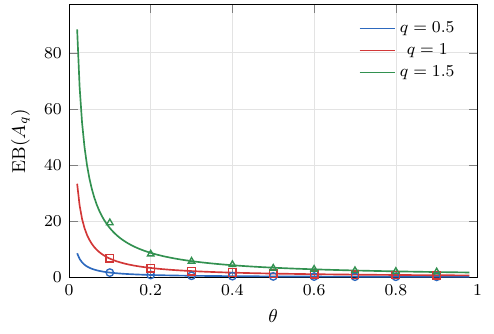}
\caption{Ergodicity breaking parameter of $A_q(t)$ as a function of $\theta$.
Solid lines denote theoretical predictions from Eq. \eqref{ebaq} for $q = 0.5$ (blue), $1.0$ (red), and $1.5$ (green). Open symbols represent the corresponding simulation results.}
\label{fig5}
\end{figure}

We now examine observables that are non-integrable with respect to the infinite invariant density, specifically power-law functionals of $X(t)$
	\begin{equation}
		A_q(t)=\int_0^t X(t')^q\,dt',
		\qquad q\geq 0.
		\label{eq:Aq}
	\end{equation}
Unlike the case $q = 1$, an analytical solution to the Feynman--Kac equation for $A_q(t)$ remains elusive for general $q$. Nonetheless, the first two moments of $A_q(t)$ can be derived using the Markov property of the Feller process. For simplicity, we set $X(0) = x_0 = 0$. The first moment is
\begin{eqnarray}
    \left\langle A_{q}(t)\right\rangle =\int^{t}_{0}dt'\left\langle X(t')^{q}\right\rangle =\int^{t}_{0}dt'\int^{\infty}_{0}x^{q}P(x,t'|0)dx,
\end{eqnarray}
where $P(x,t|0)=\lim_{x_0\to0}P(x,t|x_0) = x^{\theta-1}e^{-x/t}/[t^\theta\Gamma(\theta)]$ from Eq.~\eqref{prop2}, giving
\begin{eqnarray}
    \left\langle A_{q}(t)\right\rangle =\frac{\Gamma(q+\theta)}{\Gamma(\theta)}\frac{t^{1+q}}{1+q}.
\end{eqnarray}
The second moment is evaluated using the two-point joint PDF $P(x_2,t_2;x_1,t_1|0)$
\begin{eqnarray}
    \left\langle A_{q}(t)^{2}\right\rangle =\int^{t}_{0}dt_{2}\int^{t}_{0}dt_{1}\left\langle X(t_{2})^{q}X(t_{1})^{q}\right\rangle= 2\int^{t}_{0}dt_{2}\int^{t_{2}}_{0}dt_{1}\,\langle X(t_{1})^{q}X(t_{2})^{q}\rangle,
    \label{m2}
\end{eqnarray}
where
$$
\left\langle X(t_{2})^{q}X(t_{1})^{q}\right\rangle =\int^{\infty}_{0}x^{q}_{2}dx_{2}\int^{\infty}_{0}x^{q}_{1}dx_{1}P(x_{2},t_{2};x_{1},t_{1}|0).
$$
Using property~\eqref{mark2} alongside propagator~\eqref{prop2}, the joint PDF is expressed as
\begin{eqnarray}
    P(x_{2},t_{2};x_{1},t_{1}|0)=\frac{\left(x_{1}x_{2}\right)^{\frac{\theta-1}{2}}}{t^{\theta}_{1}(t_{2}-t_{1})\Gamma(\theta)}\exp\left(-\frac{x_{2}+x_{1}}{t_{2}-t_{1}}-\frac{x_{1}}{t_{1}}\right)I_{\theta-1}\left(\frac{2\sqrt{x_{2}x_{1}}}{t_{2}-t_{1}}\right).
\end{eqnarray}
Integrating over $x_1$ and $x_2$ yields
\begin{eqnarray*}
 \left\langle X(t_{2})^{q}X(t_{1})^{q}\right\rangle &=
 &\left[\frac{\Gamma(q+\theta)}{\Gamma(\theta)}\right]^{2}\frac{t^{q}_{1}(t_{2}-t_{1})^{\theta+2q}}{t^{q+\theta}_{2}}{}_{2}F_{1}\left(q+\theta,q+\theta;\theta;\frac{t_{1}}{t_{2}}\right)\\
 &=&\left[\frac{\Gamma(q+\theta)}{\Gamma(\theta)}\right]^{2}(t_{1}t_{2})^{q}{}_{2}F_{1}\left(-q,-q;\theta;\frac{t_{1}}{t_{2}}\right), 
\end{eqnarray*}
where we applied Euler's identity for the Gauss hypergeometric function. Substituting this into Eq.~\eqref{m2} and changing variables to $u=t_1/t_2$ gives
$$
\left\langle A_{q}(t)^{2}\right\rangle =\frac{t^{2(q+1)}}{q+1}\left[\frac{\Gamma(q+\theta)}{\Gamma(\theta)}\right]^{2}\int^{1}_{0}\,u^{q}\,{}_{2}F_{1}\left(-q,\,-q;\,\theta;\,u\right)du.
$$
Integrating the hypergeometric function via Eq.~7.512.12 of Ref.~\cite{GrRy07} yields
\begin{eqnarray}
    \left\langle A_{q}(t)^{2}\right\rangle =\frac{t^{2(q+1)}}{(q+1)^{2}}\left[\frac{\Gamma(q+\theta)}{\Gamma(\theta)}\right]^{2}{}_{3}F_{2}\left(-q,\,-q,\,q+1;\,\theta,\,q+2;\,1\right).
\end{eqnarray}
The ergodicity breaking parameter is thus
\begin{eqnarray}
    \mathrm{EB}(A_{q})=\lim_{t\to\infty}\frac{\left\langle A_{q}(t)^{2}\right\rangle }{\left\langle A_{q}(t)\right\rangle ^{2}}-1={}_{3}F_{2}\left(-q,\,-q,\,q+1;\,\theta,\,q+2;\,1\right)-1,
    \label{ebaq}
\end{eqnarray}
which clearly differs from Eq.~\eqref{ebx}, confirming that non-integrable power observables $A_q(t)$ do not obey the Darling-Kac theorem.
In Figure \ref{fig5} we plot the EB parameter of the functional $A_q(t)$ as function of $\theta$. The agreement between the theoretical results and simulations is excellent. 

\begin{figure}[htbp]
\centering
\includegraphics[width=0.65\linewidth]{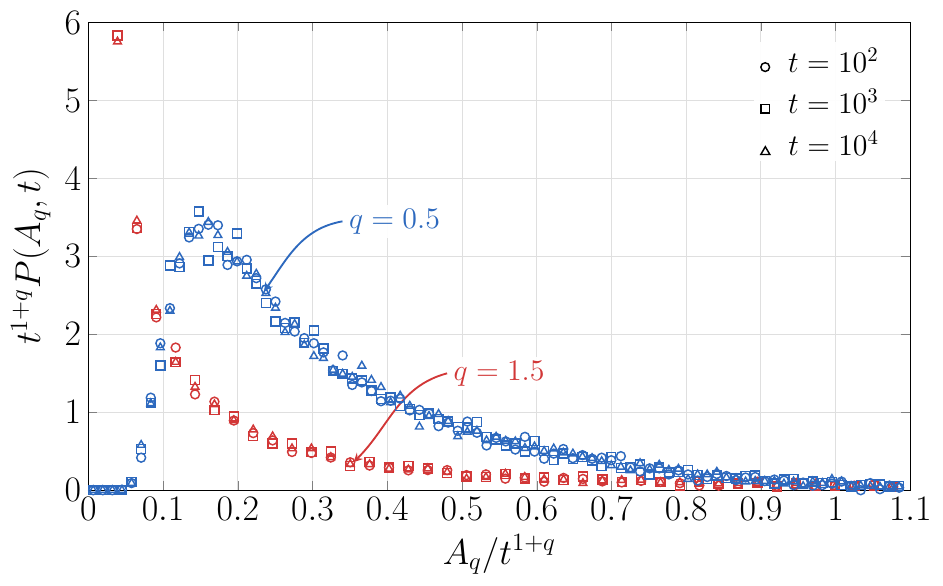}
\caption{
Data collapse of the probability density function $t^{1+q}P(A_q, t)$  vs $A_q/t^{1+q}$ for $q = 0.5$ (blue) and $q=1.5$ (red) at observation times $t = 10^2$ (circles), $10^3$ (squares), and $10^4$ (triangles).}
\label{fig6}
\end{figure}

Another key feature of $A_q(t)$ is its self-similar scaling. Scaling time by a constant factor $c$ ($t \to ct$) in Eq.~\eqref{eq:feller} and exploiting the scaling property $W(ct)\stackrel{d}{=}\sqrt{c}W(t)$, it can be shown that the unconfined Feller process satisfies $X(ct) \stackrel{d}{=} c X(t)$. This property implies the scaling law for $A_q(t)$. Substituting $u=t'/t$ into Eq.~\eqref{eq:Aq} yields
\begin{eqnarray}
    A_q(t)=\int_0^t X(t')^q\,dt'=t^{1+q}\int_0^1 X(u)^q\,du.
\end{eqnarray}
Thus, $A_q(t)$ scales as $t^{1+q} \chi$, where $\chi = \int_0^1 X(u)^q du$ is a time-independent random variable. Denoting the PDF of $\chi$ by $\mathcal{P}_q(\chi)$, the variable transformation $\chi = A_q / t^{1+q}$ establishes the scaling form of $P(A_q, t)$
\begin{eqnarray}
    P(A_{q},t)=\frac{1}{t^{1+q}}\mathcal{P}_q\left(\frac{A_{q}}{t^{1+q}}\right).
    \label{pchi}
\end{eqnarray}
In Figure \ref{fig6} check the validity of Eq. \eqref{pchi}. For fixed value of $q$ the simulation data of $t^{1+q}P(A_q,t)$ collapse on the same curve $\mathcal{P}_q(\chi)$ where $\chi=A_q/t^{1+q}$.

\section{Discussion}
\label{sec:discussion}

The non-confined Feller diffusion ($0 < \theta < 1$) serves as a foundational archetype for understanding stochastic dynamics driven by multiplicative noise in the presence of an non-normalized stationary state. Because it is not normalized, probability mass continually escapes toward infinity, placing the system firmly under the domain of infinite ergodic theory \cite{Aaronson1997, Korabel2009, Akimoto2010}. Our results provide a local-field interpretation of weak ergodicity breaking: the local-time $L(x,t)$ reveals how the non-normalizable stationary measure is reconstructed pointwise across the continuous spatial domain, while all local temporal fluctuations become fundamentally locked to a single global renewal amplitude $\zeta$.

From a mathematical standpoint, because the exact Laplace-space transition propagator scales asymptotically as $\widetilde{P}(x,s|x_0) \sim s^{-(1-\theta)} x^{\theta-1} / \Gamma(\theta)$ for small $s$, the unconfined Feller process strictly satisfies the operational conditions of the Darling--Kac theorem \cite{DarlingKac1957, Aaronson1997}. Consequently, the normalized local time universally converges to a Mittag-Leffler limit distribution, with the fluctuation index mapped directly to the physical drift parameter via $\gamma = 1 - \theta$. While the emergence of this universality class might be theoretically anticipated within the abstract framework of infinite ergodic theory, its rigorous manifestation in the Feller model is highly non-trivial. The underlying Langevin dynamics are driven by multiplicative noise and are subject to a singular boundary at the origin. A priori, it is not obvious that the delicate balance between space-dependent diffusion and zero-flux boundary conditions would preserve pure fractional Mittag-Leffler scaling without generating anomalous boundary-layer corrections or non-universal temporal cutoffs. Our exact analytical treatment using Bessel functions explicitly proves the ultimate structural robustness of this scaling.

More importantly, whereas the standard Darling--Kac theorem is typically formulated for a single localized functional evaluated at a specific spatial point, our exact multi-point calculation reveals a far deeper physical architecture. Evaluating the normalized two-point spatial cross-correlation function $R(x_1, x_2, t)$
we demonstrate the exact mathematical cancellation of the spatial geometry in the long-time limit $t \to \infty$, yielding a level-independent universal plateau.
This exact cancellation proves the asymptotic emergence of rigid, global spatial coherence across the entire state space. It establishes that Mittag-Leffler statistics do not arise as an uncoupled collection of independent local fluctuations at different positions $x_1$ and $x_2$. Instead, the local-time field $L(x,t)$ undergoes an exact asymptotic spatiotemporal factorization $L(x,t)\underset{t\to\infty}{\approx}\mathcal{I}(x)\Phi(t)$
where $\Phi (t)\sim\zeta$ acts as a singular, space-independent stochastic amplitude, distributed according to $\mathcal{M}_{1-\theta}(\zeta)$,—a universal "sample-path clock"—that synchronously modulates the occupation density across all spatial points simultaneously.

From a physical standpoint, the level-independence of $R(x_1, x_2, t)$ and the underlying factorization  reflect a synchronized mode of spatial exploration governed by global excursion loops. 
Finally, this local-field framework establishes a sharp physical taxonomy when contrasting integrable versus non-integrable observables. Occupation time $T_a(t)$ and local time $L(x,t)$ belong to $L^1(\mathcal{I})$, as their underlying spatial weight functions are integrable against the invariant density $\mathcal{I}(x) = x^{\theta-1}$. As a direct consequence, they inherit the universal Mittag-Leffler scaling dictated by the global amplitude $\zeta$. Conversely, for non-integrable power-law observables $A_q(t)$ ($q > 0$), the observable $U[x] = x^q$ diverges when integrated against $\mathcal{I}(x)$. In this non-integrable regime, extreme excursions far from the origin dominate the path integral, breaking Darling--Kac universality. The process escapes the $L^1(\mathcal{I})$ framework.

\section{Conclusions}
\label{sec:conclusions}

We have developed a comprehensive local-time perspective to unravel infinite ergodic behavior, multi-point correlation structures, and non-equilibrium fluctuation statistics in non-confined Feller process. The central result of this work is the analytical proof that the local-time field undergoes an asymptotic spatiotemporal factorization into the infinite invariant density and a space-independent random amplitude $\Phi (t)$. This formulation offers a unified framework that seamlessly connects local integrable occupation observables to global non-integrable trajectory functionals in multiplicative stochastic diffusions.

The principal physical takeaway is that the non-confined Feller process exhibits rigid global spatial coherence. In the vast majority of spatial random fields, turbulent media, or standard diffusive systems, two-point correlation functions depend explicitly on spatial distance $|x_1 - x_2|$ and decay systematically as the separation between observation points increases. In stark contrast, for the non-confined Feller diffusion, spatial distance effectively disappears from the asymptotic correlation structure which converges to a strictly level-independent universal constant $2[\Gamma(2-\theta)]^2/\Gamma(3-2\theta)$. Whether two observation levels are infinitesimally close or separated by very large distances, their normalized cross-correlation remains strictly identical.

The level-independent local-time correlation extends the usual one-observable Darling--Kac picture to the spatial structure of the occupation field. Similar infinite-density phenomenology occurs in dissipative optical-lattice models of cold atoms \cite{Afek2023}. However, the present result concerns the joint local-time field: normalized occupation densities at distinct fixed levels become asymptotically locked in mean square. Establishing whether the same multipoint structure occurs in logarithmic optical-lattice dynamics remains an open question.
	
	\begin{acknowledgments}
		The author acknowledge the financial support of the Ministerio de Ciencia e Innovaci\'on (Spanish government) under
Grant No. PID2021-122893NB-C22.
	\end{acknowledgments}

	\appendix
\section{Numerical methods}
\label{app:numerics}

All numerical results were obtained from independent Monte Carlo
realizations of Eq. \eqref{eq:feller_confined2}. Two numerical procedures were used. Local-time and occupation-time
observables were evaluated on trajectories generated with a
positivity-preserving full-truncation Euler scheme,
\begin{equation}
X_{n+1}
=
\max\left[
X_n+\theta\Delta t
+\sqrt{2\max(X_n,0)}\,\Delta W_n,0
\right],
\label{eq:app_euler}
\end{equation}
where $\Delta W_n$ are independent normal variables with variance
$\Delta t$. The time step was varied between $10^{-3}$ and $10^{-2}$
to check discretization effects. The local time was regularized by a box kernel,
\begin{equation}
L_{\epsilon}^{(k)}(x,t)
=
\frac{\Delta t}{2\epsilon}
\sum_{n=1}^{\lfloor t/\Delta t\rfloor}
\mathbb{I}\left(|X_n^{(k)}-x|\leq\epsilon\right),
\label{eq:app_local_time}
\end{equation}
with $\epsilon=0.05$--$0.08$. $\mathbb{I}(\cdot)$ is the indicator function, $(k)$ stand for the $k$-th trajectory and $\lfloor \cdot\rfloor$ is the floor function. The  occupation time was computed
from
\begin{equation}
T_a^{(k)}(t)
=
\Delta t
\sum_{n=1}^{\lfloor t/\Delta t\rfloor}
\mathbb{I}\left(X_n^{(k)}\leq a\right),
\qquad
T_a^{+(k)}(t)=t-T_a^{(k)}(t).
\label{eq:app_occupation}
\end{equation}
The local-time correlations and EB parameters were estimated directly
from sample moments. Ensembles between $10^4$ and $10^6$ realizations
were used, depending on the observable and the tail sensitivity of the
quantity being estimated. The occupation-time calculations used
$a=1$ and observation times $t=10^2$, $10^3$, and $10^4$.

For the density collapse of $A_q(t)$, the process was
sampled at the grid points from its exact finite-time transition law.
Conditional on $X_n$, one has
\begin{equation}
X_{n+1}
\overset{d}{=}
\frac{\Delta t}{2}
\chi'^2_{2\theta}
\left(\frac{2X_n}{\Delta t}\right),
\label{eq:app_exact_transition}
\end{equation}
where $\chi'^2_{\nu}(\Lambda)$ is a noncentral chi-squared random
variable with $\nu$ degrees of freedom and noncentrality parameter 
$\Lambda$ \cite{RevuzYor1999,MalhamWiese2013}. This update preserves non-negativity and introduces no
time-stepping error in the sampled state values. The functional $A_q(t)$ was evaluated by the trapezoidal rule,
\begin{equation}
A_q^{(k)}(t)
\simeq
\frac{\Delta t}{2}
\sum_{n=0}^{N_{\mathrm{step}}-1}
\left[
(X_n^{(k)})^q+(X_{n+1}^{(k)})^q
\right].
\label{eq:app_Aq_quadrature}
\end{equation}
where $N_{\mathrm{step}}=t/\Delta t$. For each pair $(q,t)$ in Fig.~\ref{fig6}, we generated
$N_{\mathrm{traj}}=1.2\times10^4$ trajectories with
$N_{\mathrm{step}}=1800$ intervals. The calculations used $\theta=0.4$,
$q=0.5$ and $1.5$, and $t=10^2$, $10^3$, and $10^4$.

The density of $Y_q=A_q/t^{1+q}$ was estimated with normalized
histograms. For each $q$, a common set of 85 equally spaced bins was
used for all three observation times. The bin edges were determined
from the pooled samples, with the upper boundary fixed at the $99.5$th
percentile to prevent a few extreme trajectories from setting the
entire displayed range. This choice affects only the histogram window
and does not modify the simulated samples. The figure shows
$0\leq A_q/t^{1+q}\leq1.1$ and
$0\leq t^{1+q}P(A_q,t)\leq6$. Numerical convergence was checked by
varying the time resolution and the number of histogram bins.

\bibliography{main}

@book{Go20,
	author = {Gorenflo, Rudolf and Kilbas, Anatoly A. and Mainardi, Francesco and Rogosin, Sergei},
	doi = {10.1007/978-3-662-61550-8},
	isbn = {9783662615508},
	issn = {2196-9922},
	journal = {Springer Monographs in Mathematics},
	publisher = {Springer Berlin Heidelberg},
	title = {Mittag-Leffler Functions, Related Topics and Applications},
	url = {http://dx.doi.org/10.1007/978-3-662-61550-8},
	year = {2020}}

@article{Ba01,
  title = {Fractional Fokker-Planck equation, solution, and application},
  author = {Barkai, E.},
  journal = {Physical Review E},
  volume = {63},
  issue = {4},
  pages = {046118},
  numpages = {17},
  year = {2001},
  month = {Mar},
  publisher = {American Physical Society},
  doi = {10.1103/PhysRevE.63.046118},
  url = {https://link.aps.org/doi/10.1103/PhysRevE.63.046118}
}

@article{Penson2010,
  title = {Exact and Explicit Probability Densities for One-Sided L\'evy Stable Distributions},
  author = {Penson, K. A. and G\'orska, K.},
  journal = {Physical Review Letters},
  volume = {105},
  issue = {21},
  pages = {210604},
  numpages = {4},
  year = {2010},
  month = {Nov},
  publisher = {American Physical Society},
  doi = {10.1103/PhysRevLett.105.210604},
  url = {https://link.aps.org/doi/10.1103/PhysRevLett.105.210604}
}

@article{Feller1951,
  author  = {Feller, W.},
  title   = {Two singular diffusion problems},
  journal = {Annals of Mathematics},
  volume  = {54},
  pages   = {173--182},
  year    = {1951}
}

@article{MasoliverPerello2012,
  author  = {Masoliver, J. and Perell{\'o}, J.},
  title   = {First-passage and escape problems in the {Feller} process},
  journal = {Physical Review E},
  volume  = {86},
  number  = {4},
  pages   = {041116},
  year    = {2012}
}

@book{RevuzYor1999,
  author    = {Revuz, D. and Yor, M.},
  title     = {Continuous Martingales and Brownian Motion},
  publisher = {Springer},
  address   = {Berlin},
  year      = {1999}
}

@book{Aaronson1997,
  author    = {Aaronson, J.},
  title     = {An Introduction to Infinite Ergodic Theory},
  publisher = {American Mathematical Society},
  address   = {Providence},
  year      = {1997}
}

@article{Ke78,
  author  = {Kent, John},
  title   = {Some probabilistic properties of {Bessel} processes},
  journal = {The Annals of Probability},
  volume  = {6},
  number  = {5},
  pages   = {760--770},
  year    = {1978},
  doi     = {10.1214/aop/1176995426}
}

@article{MaPe12,
  author  = {Masoliver, Jaume and Perell{\'o}, Josep},
  title   = {First-passage and escape problems in the {Feller} process},
  journal = {Physical Review E},
  volume  = {86},
  number  = {4},
  pages   = {041116},
  year    = {2012},
  doi     = {10.1103/PhysRevE.86.041116}
}

@article{GoYo03,
  author  = {G{\"o}ing-Jaeschke, Anja and Yor, Marc},
  title   = {A survey and some new results on hitting times for {Bessel} processes},
  journal = {Finance and Stochastics},
  volume  = {7},
  number  = {2},
  pages   = {253--275},
  year    = {2003},
  doi     = {10.1007/s007800200083}
}

@book{GrRy07,
  title     = {Table of Integrals, Series, and Products},
  author    = {Gradshteyn, Izrail Solomonovich and Ryzhik, Iosif Moiseevich},
  editor    = {Jeffrey, Alan and Zwillinger, Daniel},
  edition   = {7th},
  publisher = {Academic Press},
  address   = {Amsterdam},
  year      = {2007},
  isbn      = {978-0-12-373637-6}
}

@article{Bouchaud1990,
  title = {Anomalous diffusion in disordered media: statistical mechanisms, models and physical applications},
  author = {Bouchaud, Jean-Philippe and Georges, Antoine},
  journal = {Physics Reports},
  volume = {195},
  number = {4-5},
  pages = {127--293},
  year = {1990},
  publisher = {Elsevier}
}

@article{Cox1985,
  title = {A theory of the term structure of interest rates},
  author = {Cox, John C and Ingersoll Jr, Jonathan E and Ross, Stephen A},
  journal = {Econometrica},
  volume = {53},
  number = {2},
  pages = {385--407},
  year = {1985}
}

@book{Gardin2009,
  title = {Stochastic Methods: A Handbook for the Natural and Social Sciences},
  author = {Gardiner, Crispin},
  year = {2009},
  publisher = {Springer},
  address = {Berlin}
}

@book{Horsthemke1984,
  title = {Noise-Induced Transitions: Theory and Applications in Physics, Chemistry, and Biology},
  author = {Horsthemke, Werner and Lefever, Ren{\'e}},
  year = {1984},
  publisher = {Springer-Verlag},
  address = {Berlin}
}

@book{vanKampen2007,
  title = {Stochastic Processes in Physics and Chemistry},
  author = {van Kampen, N. G.},
  year = {2007},
  publisher = {Elsevier},
  address = {Amsterdam}
}

@article{Schenzle1979,
  title = {Multiplicative stochastic processes in statistical physics},
  author = {Schenzle, Axel and Brand, Helmut},
  journal = {Physical Review A},
  volume = {20},
  number = {4},
  pages = {1628--1647},
  year = {1979}
}

@article{Barkai2014,
  title = {Strange kinetics of single molecules in living cells},
  author = {Barkai, Eli and Garini, Yuval and Metzler, Ralf},
  journal = {Physics Today},
  volume = {65},
  number = {8},
  pages = {29--35},
  year = {2012}
}

@article{Birkhoff1931,
  title = {Proof of the ergodic theorem},
  author = {Birkhoff, George D.},
  journal = {Proceedings of the National Academy of Sciences of the United States of America},
  volume = {17},
  number = {12},
  pages = {656--660},
  year = {1931}
}

@book{Risken1989,
  title = {The Fokker-Planck Equation: Methods of Solution and Applications},
  author = {Risken, Hannes},
  year = {1989},
  publisher = {Springer},
  address = {Berlin}
}

@article{Korabel2009,
  title = {Pesin-Type Identity along the Transition to Anomalous Diffusion},
  author = {Korabel, Nickolay and Barkai, Eli},
  journal = {Physical Review Letters},
  volume = {102},
  number = {5},
  pages = {050601},
  year = {2009}
}

@article{Akimoto2010,
  title = {Role of infinite invariant measure in anomalous diffusion},
  author = {Akimoto, Takuma and Miyaguchi, Tomoshige},
  journal = {Physical Review E},
  volume = {82},
  number = {3},
  pages = {030102(R)},
  year = {2010}
}

@article{Rebenshtok2007,
  title = {Distribution of Time-Averaged Observables for Weak Ergodicity Breaking},
  author = {Rebenshtok, Alon and Barkai, Eli},
  journal = {Physical Review Letters},
  volume = {99},
  number = {21},
  pages = {210601},
  year = {2007}
}

@article{Deng2009,
  title = {Ergodic properties of fractional Brownian motion},
  author = {Deng, Weihua and Barkai, Eli},
  journal = {Physical Review E},
  volume = {79},
  number = {1},
  pages = {011112},
  year = {2009}
}

@article{DarlingKac1957,
  title = {On occupation times for Markov processes},
  author = {Darling, D. A. and Kac, M.},
  journal = {Transactions of the American Mathematical Society},
  volume = {84},
  number = {2},
  pages = {444--458},
  year = {1957}
}

@article{He2008,
  title = {Random Walks, Anomalous Diffusion, and Ergodicity Breaking},
  author = {He, Y. and Burov, S. and Metzler, R. and Barkai, E.},
  journal = {Physical Review Letters},
  volume = {101},
  number = {5},
  pages = {058101},
  year = {2008}
}

@article{Afek2023,
  author  = {G. Afek and N. Davidson and D. A. Kessler and E. Barkai},
  title   = {Colloquium: Anomalous statistics of laser-cooled atoms
             in dissipative optical lattices},
  journal = {Rev. Mod. Phys.},
  volume  = {95},
  pages   = {031003},
  year    = {2023},
  doi     = {10.1103/RevModPhys.95.031003}
}

@article{MalhamWiese2013,
  author  = {S. J. A. Malham and A. Wiese},
  title   = {Chi-square simulation of the {CIR} process and the
             {Heston} model},
  journal = {Int. J. Theor. Appl. Finance},
  volume  = {16},
  number  = {3},
  pages   = {1350014},
  year    = {2013},
  doi     = {10.1142/S0219024913500143}
}

\end{document}